\documentclass[aps,prl,reprint,showpacs,superscriptaddress]{revtex4-1}
\pdfoutput=1
\usepackage{graphicx}

\begin{document}

\title{Adaptive Synchronization and Anticipatory Dynamical System}

\author{Ying-Jen Yang}

\affiliation{Institute of Physics, Academia Sinica, Taipei, Taiwan 115, Republic
of China}

\affiliation{Department of Physics, National Taiwan University, Taipei, Taiwan,
Republic of China}

\author{Chun-Chung Chen}

\email{cjj@phys.sinica.edu.tw}

\affiliation{Institute of Physics, Academia Sinica, Taipei, Taiwan 115, Republic
of China}

\author{Pik-Yin Lai}

\affiliation{Institute of Physics, Academia Sinica, Taipei, Taiwan 115, Republic
of China}

\affiliation{Department of Physics and Center for Complex Systems, National Central
University, Chungli, Taiwan 320, Republic of China}

\author{C. K. Chan}

\affiliation{Institute of Physics, Academia Sinica, Taipei, Taiwan 115, Republic
of China}

\affiliation{Department of Physics and Center for Complex Systems, National Central
University, Chungli, Taiwan 320, Republic of China}

\date{July 19, 2015}
\begin{abstract}
Many biological systems can sense periodical variations in a stimulus
input and produce well-timed, anticipatory responses after the input
is removed. Such systems show memory effects for retaining timing
information in the stimulus and cannot be understood from traditional
synchronization consideration of passive oscillatory systems. To understand
this anticipatory phenomena, we consider oscillators built from excitable
systems with the addition of an adaptive dynamics. With such systems,
well-timed post-stimulus responses similar to those from experiments
can be obtained. Furthermore, a well-known model of working memory
is shown to possess similar anticipatory dynamics when the adaptive
mechanism is identified with synaptic facilitation. The last finding
suggests that this type of oscillators can be common in neuronal systems
with plasticity.
\end{abstract}


\maketitle
The interaction between a dynamical system capable of oscillatory
behavior and an external periodic stimulus input represents the fundamental
physics underlying many biological and engineering systems \cite{pikovsky_synchronization:_2003,gros_complex_2011}.
Under the titles of synchronization or entrainment, active studies
have been devoted to various aspects of these systems, for example,
the existence and degree of synchrony under different parameter conditions,
the stability of the synchronous state, the influence of noise \cite{pikovsky_synchronization:_2003},
and, recently, the efficiency of entrainment \cite{zlotnik_optimal_2013}.
While these mostly involve how the systems enter or stay under the
rhythmic interaction, the post-interaction behavior of the systems
often plays important roles in many biological functions \cite{winfree_geometry_2001}.
This potentially allows the biological systems to recognize the temporal
patterns in the environment and produce anticipatory responses to
help their survival. Examples of such transient responses following
the end of stimulus have been observed in ganglion cells of retina
under light stimulus \cite{schwartz_detection_2007}, growth of slime
mold influenced by varying humidity \cite{saigusa_amoebae_2008},
and the optic tectum of zebrafish conditioned with moving periodic
scenery \cite{sumbre_entrained_2008}. However, the underlying physics
for this time perception mechanism is still unclear and there is a
ongoing debate on whether a clock is needed \cite{bueti_encoding_2010}. 

To explain the anticipative dynamics, it has been proposed that the
periods of various lengths can be built into the structure of a network
with loops of various sizes \cite{mi_long-period_2013}. External
stimulations of a particular period will then activate a particular
circuit in the system to provide the memory effect. This implementation
of ``memory'' with pre-fabricated structure is similar in spirit
to the phase model used to understand the anticipation of slime mold
\cite{saigusa_amoebae_2008}. In this phase model, oscillators with
various periods must be first put into the system before an external
stimulation of a particular frequency can be used to entrain or synchronize
the oscillators with the same frequency to provide the anticipation
effect. An obvious objection to this type of models of pre-fabricated
structures is that a continuous change in external periods will not
produce a continuous change in the response; contrary to findings
of experiments with retina.

In our view, the anticipation effect is the adaptation of a biological
system to a periodic stimulation. In the following, we show the function
of such anticipation can be achieved by a minimal dynamical system
with two-dimensional oscillations controlled by a slow adaptive dynamics.
A surprisingly well retention of periodicity information for a range
of stimulus period can be obtained with single parameter adjustment,
and thus can be easily targeted evolutionarily. Such dynamical mechanism
does not necessarily describe a dominating microscopic cellular process
or pathway. It can also act as a coarse-grained, thermodynamic description
for systems of many degrees of freedom. In a neural network model
for working memory \cite{mongillo_synaptic_2008}, we show an anticipative
dynamics can be produced in the mean-field level of the network with
the adaptation coming from the residual calcium dynamics of synaptic
facilitation. This suggests a validity test for our idea in experiments.
Furthermore, our model demonstrates that the anticipative mechanism
can be quite generic and may be wide spread in natural neural systems;
hinting that the perception of time in biological systems does not
require the existence of a clock.

To substantiate our considerations, we use a reduced FitzHugh-Nagumo
(FHN) model \cite{fitzhugh_impulses_1961} defined by the equations,
\begin{equation}
\frac{dv}{dt}=v-\frac{v^{3}}{3}-w+I_{\text{ext}}\left(t\right),\label{eq:dynamics of v}
\end{equation}
\begin{equation}
\frac{dw}{dt}=\frac{1}{\tau}\left(v+a\right).\label{eq:dynamics of w}
\end{equation}
Note that we are just using the generic excitable properties of the
model to demonstrate our idea. The excited state of the FHN model
considered here does not represent an action potential and the time
scale of the system can be much longer than the spiking dynamics typically
associated with the FHN model. Usually, the system parameter $a$
is a constant for the FHN model. However, to perceive the period of
a periodic stimulus input, we turn $a$ into a dynamic variable, allowing
the system to adapt to different limiting behavior. It is known that
the dynamics of $v$ and $w$ can be synchronized to that of the periodic
$I_{\text{ext}}$ when $a$ and $\tau$ are properly chosen. The adaptation
of $a$ to the external period can be written in general as: 
\begin{equation}
\frac{da}{dt}=\frac{1}{\tau_{a}}\left(\hat{a}-a\right)\label{eq:relaxing a dynamics}
\end{equation}
where $\hat{a}\equiv g\left(v,w\right)$ is the entrained value of
$a$ and we assume that the adaptive process is characterized by a
single time scale of relaxation much slower than the FHN dynamics
($\tau_{a}\gg\tau$). The physical meaning of such requirement is
that the value of $a$ will be close to the entrained value $\hat{a}$
when there is external stimulation and will relax back to its resting
value $a_{0}$ when the stimulation is removed. Anticipative dynamics
will be manifested during the relaxation of $a$ from $\hat{a}$ to
$a_{0}$. Our task is to find a reasonable form for $g\left(v,w\right)$
to ensure the adaptation dynamics is stable.

As adaptation is a slower process when compared to the oscillatory
dynamics of the system, we expect $\left\langle \hat{a}\right\rangle $
to depend on moments of $v$ and $w$, where $\left\langle \cdot\right\rangle $
indicates time-averaged values over stationary cycles of the driven
system. It can be shown that since $\left\langle v\right\rangle =-\left\langle a\right\rangle $
and $\left\langle w\right\rangle =\left\langle v\right\rangle -\frac{1}{3}\left\langle v^{3}\right\rangle +\left\langle I_{\text{ext}}\left(t\right)\right\rangle $,
only $\left\langle w\right\rangle $ contains significant information
of $I_{\text{ext}}$. Thus, to the lowest order, the function $g\left(v,w\right)$
is of the form 
\begin{equation}
g=a_{c}-pw\label{eq:simplest}
\end{equation}
where $a_{c}$ and $p$ are constant parameters that need to be tuned
for the best behavior. To maintain a stable fixed point at $a=a_{0}>1$
under the adaptation dynamic (\ref{eq:relaxing a dynamics}), in the
absence of external stimulus, the parameters in Eq. (\ref{eq:simplest})
should satisfy \footnote{Please see Supplementary Materials at {[}PRE website{]} for derivation
of relation (\ref{eq:ac-in-a0}) and the parameter dependency of OSR
timing.} 
\begin{equation}
a_{c}=\left(1-p\right)a_{0}+p\frac{a_{0}^{3}}{3}.\label{eq:ac-in-a0}
\end{equation}
For a more intuitive view of the phase space, we will use $p$ and
$a_{0}$ as the control parameters for the adaptive dynamics.

\begin{figure}
\begin{raggedright}
A
\par\end{raggedright}

\begin{centering}
\includegraphics[width=8cm]{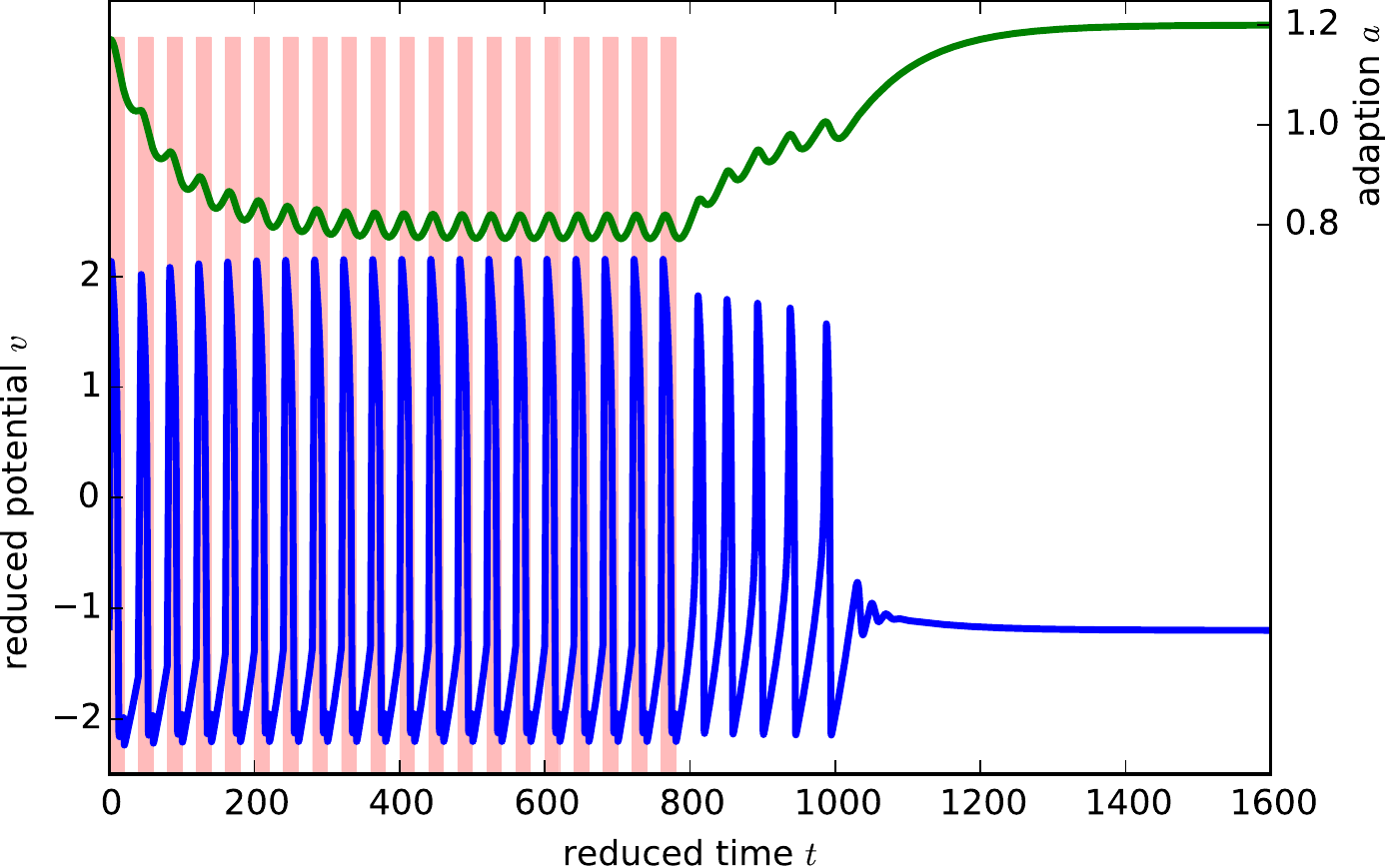}
\par\end{centering}

\begin{raggedright}
B
\par\end{raggedright}

\begin{centering}
\includegraphics[width=6cm]{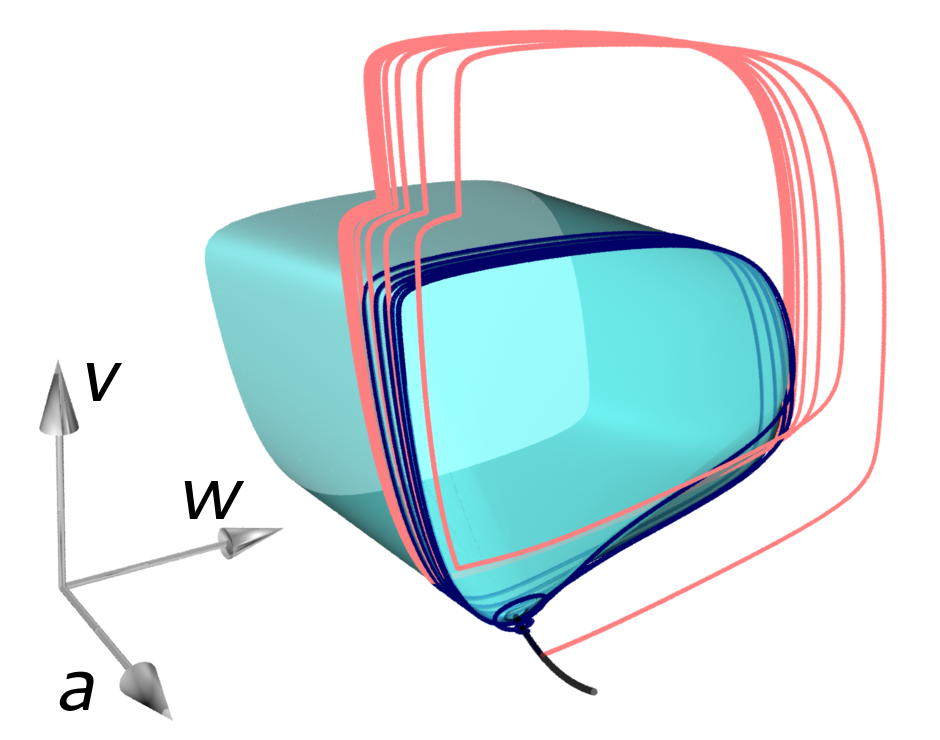}
\par\end{centering}

\raggedright{}\protect\caption{(Color online) A. Sustained oscillations of a stimulus-induced, adaptive
oscillator. The square-wave stimulus (shaded area) lasting for 20
cycles has an amplitude of 1, periods of 40, and duty ratio of $1/2$.
The control parameters for the adaptive dynamics are $a_{0}=1.2$
and $p=0.3$. B. State space trajectory for the same process when
the system is driven from the fixed point (red/light curve) and when
it relaxes back to the fixed point (blue/dark curve). The attractor
manifold of the system (cyan/shaded surface) is constructed from the
limit cycles of the simplified FitzHugh-Nagumo for fixed $a<1$ values.
The black line attached to the manifold consists of the fixed points
for $a\geq1$. \label{fig:Sustained-oscillation-from}}
\end{figure}
 Figure \ref{fig:Sustained-oscillation-from} shows the response of
our adaptive model to a transient periodic stimulation ($T_{s}=40$)
with $p=0.3$, $a_{0}=1.2$. Here, for convenience, we have used a
dimensionless time with constants $\tau=10$, and $\tau_{a}=100$.
It can be seen in Fig. \ref{fig:Sustained-oscillation-from}A that
the parameter $a$ starts from its resting value of 1.2 and decreases
to a mean value of $\left\langle \hat{a}\right\rangle \approx0.8$
as the stimulus $I_{\text{ext}}$ is applied and entrains the dynamics
of $v$ and $w$ (not shown). After $I_{\text{ext}}$ has been switched
off, $a$ relaxes back to its original value $a_{0}$. During the
relaxation period, the system produces a few residual oscillations
with periods close to that of the stimulus before returning to the
resting state. This represents the omitted stimulus responses (OSR)
\cite{schwartz_detection_2007} of the system, showing anticipative
dynamics similar to the observations in, \emph{e.g.}, zebrafish \cite{sumbre_entrained_2008}.

\begin{figure}
\begin{centering}
\includegraphics[width=8cm]{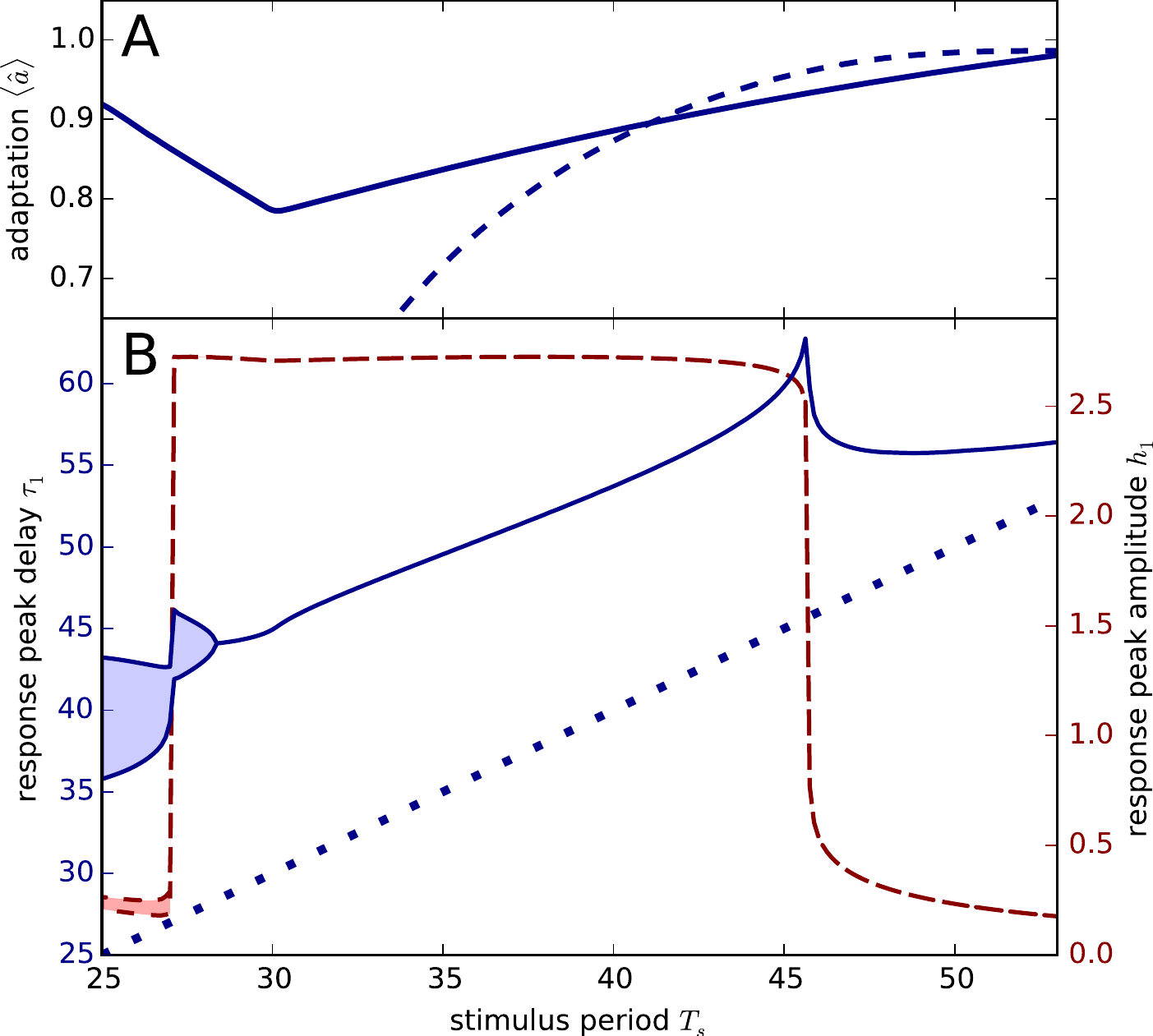}
\par\end{centering}

\protect\caption{(Color online) A.~The mean value of the adaptive variable $\left\langle \hat{a}\right\rangle $
under periodic stimulus as a function of the stimulus period (solid
line) compared with the value of $a$ in a non-adaptive FHN oscillator
that satisfies $T_{\text{fn}}\left(a\right)=T_{s}$ (dashed line).
B.~Delay time of first omitted stimulus response peak $\tau_{1}$
(solid line, left scale) as a function of stimulus period compared
to $\tau_{1}=T_{s}$ (dotted line). The dashed curve (right scale)
shows the amplitude of the response peak. The curves split into shaded
area when the driven system does not settle into a simple limit cycle.
The parameters for the adaptive dynamics are $a_{0}=1.4$ and $p=0.4$.\label{fig:mean-a-driven-and-Timing-of-first-OSR}}
\end{figure}
Simulations similar to that shown in Fig.~\ref{fig:Sustained-oscillation-from}
have also been performed for various stimulation periods. As expected,
the $\left\langle \hat{a}\right\rangle $ in the synchronized state
is a function of stimulation period (see Fig.~\ref{fig:mean-a-driven-and-Timing-of-first-OSR}A).
That is, the information of the stimulation period can be stored in
$\left\langle \hat{a}\right\rangle $. With the stored synchronized
state $a=\left\langle \hat{a}\right\rangle $, the anticipative dynamics
or OSR can be seen when $a$ relaxes back to $a_{0}$ after the external
stimulation has been removed. In a sense, the OSR or anticipative
dynamics is the byproduct of the adaptation. 

Ideally, one expect the sustained oscillation periods will match the
period of the stimulus input $T_{s}$, the position $a\approx\left\langle \hat{a}\right\rangle $
of the driven limit cycle should satisfy 
\begin{equation}
T_{\text{fn}}\left(\left\langle \hat{a}\right\rangle \right)\approx T_{s}.\label{eq:limit-cycle-position-1}
\end{equation}
where $T_{\text{fn}}\left(a\right)$ denotes the oscillation periods
in the original FHN system with a static $a$. The inverse $T_{\text{fn}}^{-1}\left(T_{s}\right)$
is shown as the dashed line in Fig.~\ref{fig:mean-a-driven-and-Timing-of-first-OSR}A.
The value of $\left\langle \hat{a}\right\rangle $, which is a function
of the stimulus period, is plotted as solid line in Fig.~\ref{fig:mean-a-driven-and-Timing-of-first-OSR}.
The condition (\ref{eq:limit-cycle-position-1}) implies two of the
curves should coincide with each other. However, we see significant
deviation even in the region where the first OSR retains the stimulus
period well (see Fig.~\ref{fig:mean-a-driven-and-Timing-of-first-OSR}B).
This suggests that other variables additional to $a$, such as the
phase information in cycle of $v$-$w$ oscillations, are also contributing
to the encoding of the periodicity information. Since we are interested
in the general phenomena of a functioning anticipation, we will focus
on the role of a single adaptive parameter for the current study.
Additionally, one can constrain the dynamics of $a$ so that condition
(\ref{eq:limit-cycle-position-1}) is satisfied, or so that one would
have $\bar{g}=T_{\text{fn}}^{-1}\left(T_{s}\right)$ for the average
of (\ref{eq:simplest}). This likely involves optimizations specific
to the systems of interests and can be a subject of future studies. 

Following \cite{schwartz_detection_2007}, the effectiveness of rhythmic
memory can be assessed by the timing of the first OSR peak right after
the stop of the periodic stimulus. Typically, the retention of the
information about stimulus period requires a one-to-one mapping of
the period to the latency. As shown in Fig.~\ref{fig:mean-a-driven-and-Timing-of-first-OSR}B,
the linear relation of the first OSR time with the stimulus period
at $p=0.4$ and $a_{0}=1.4$ shows a good retention of the stimulus-period
information: Comparing with the stimulus period, the latency $\tau_{1}$
has a constant delay of 15 units for the range of stimulus period
from 28 to 45. Such a delay depends on the details of the model and
can be interpreted as the time taken by the system to respond to the
missing stimulus. Well-timed OSR similar to Fig.~\ref{fig:mean-a-driven-and-Timing-of-first-OSR}B
can also be obtained for, say, a different value of $p$ or $\tau_{a}$
when the value of $a_{0}$ is properly chosen \cite{Note1}. As expected
from the finite oscillation range of our reduced FitzHugh-Nagumo model,
the amplitude of OSR peak drops significantly outside this range.
For a fast stimulus (small $T_{s}$ as indicated by the shaded area
in Fig.~\ref{fig:mean-a-driven-and-Timing-of-first-OSR}B), the system
may not be able to settle into a simple limit cycle with the same
small period, leading to a more complex behavior similar to the observation
in \cite{cortes_short-term_2013}. While, for a slow stimulus, the
system will return to the quiescent fixed point before producing an
OSR.

\begin{figure}
\begin{centering}
\includegraphics[width=7cm]{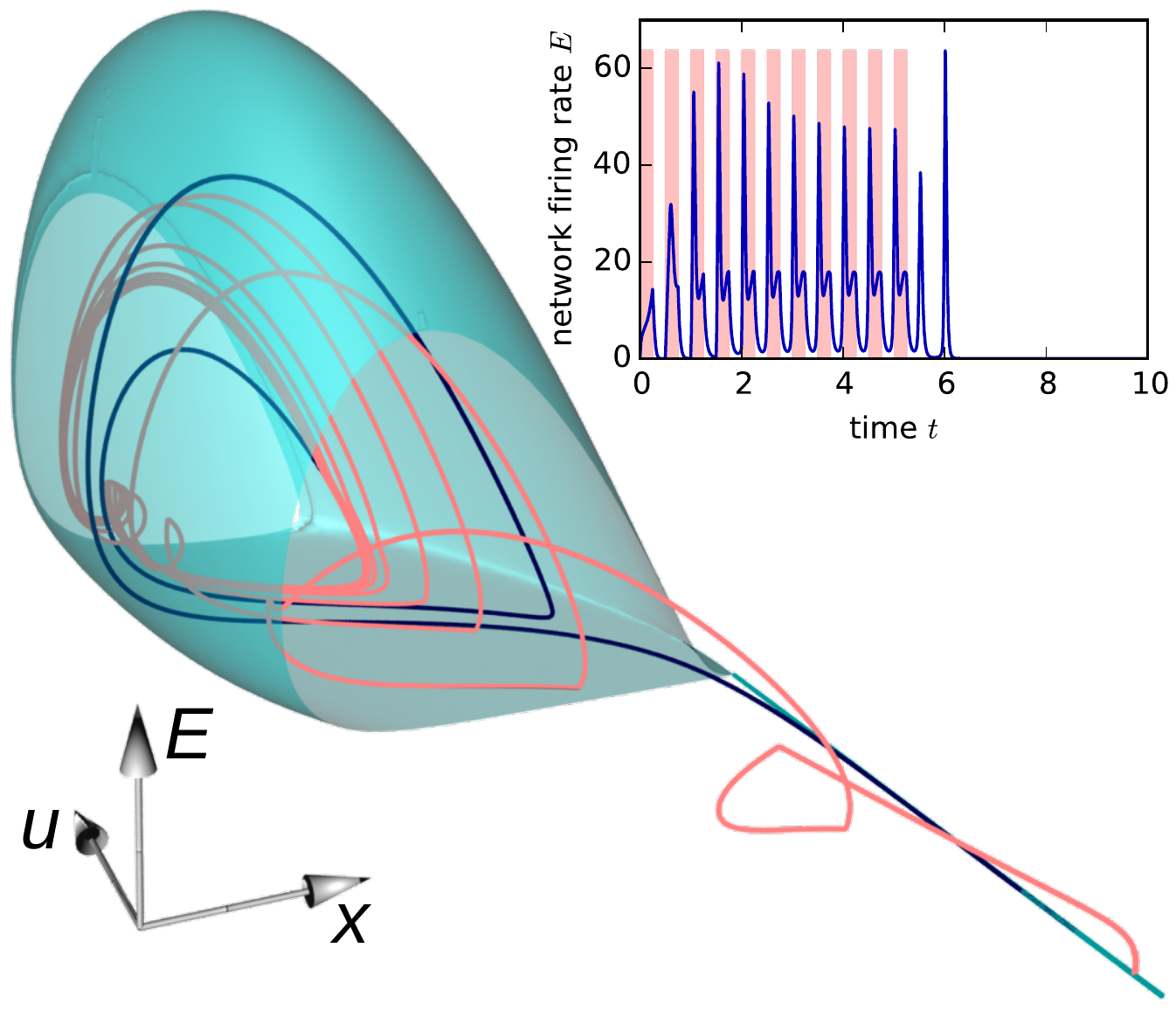}
\par\end{centering}

\protect\caption{(Color online) Anticipative behavior of mean-field dynamics for a
network model of working memory \cite{mongillo_synaptic_2008} (see
\cite{Note1}). The trajectory and attractor manifold are similar
to that described in Fig.~\ref{fig:Sustained-oscillation-from}B.
The inset shows time course of the network firing rate $E$ driven
by a square-wave stimulus (shaded) and its residual oscillations after
the stimulus stops.\label{fig:Mongillo's-mean-field}}
\end{figure}

In our conceptual model, it should be clear that all we need to implement
anticipative dynamics in an excitable system is an adaptive excitability.
Interestingly, such a mechanism has already been used in the modeling
of working memory (WM). Mongillo \emph{et al.} \cite{mongillo_synaptic_2008}
consider WM as the result of the interactions between synaptic facilitation
and depression. As illustrated in Fig.~\ref{fig:Mongillo's-mean-field},
anticipative dynamics can also be observed in such WM model \footnote{To have a quiescent state in the absence of external input, we use
a slightly modified gain function $g\left(z\right)=\alpha\ln\left[\left(1+e^{z/\alpha}\right)/2\right]$
from the Supporting Online Material of \cite{mongillo_synaptic_2008}.}. In a mean-field approximation \cite{wilson_excitatory_1972}, the
neocortical network in \cite{mongillo_synaptic_2008} is described
by the firing rate $E$ of the neural tissue, the available neurotransmitter
fraction $x$, and the probability of synaptic release $u$ (see supplementary
material of \cite{mongillo_synaptic_2008}). The parameter $u$ there
plays a similar role as $a$ in our FHN model, namely, the value of
$u$ determines whether the stationary state of the system is at a
stable fixed point or on a limit cycles. Furthermore, the adaptation
dynamic of $u$ comes from synaptic facilitation \cite{tsodyks_neural_1998,mongillo_synaptic_2008}
and relates to the calcium concentration of the presynaptic cell.
One interpretation of such relation is that the perception of time
is being stored in the calcium concentration in these cells.

In a broader view of these systems, the transient, anticipative oscillations
following the end of the periodic stimulus are similar to the burster
dynamics found in neurosciences \cite{izhikevich_dynamical_2006}
where a slow variable drive the system in and out of an oscillatory
region of the state space. Since the adaptive dynamics is much slower
than the oscillation, the trajectories of these systems mostly follow
their attractor manifolds as shown in Figs. \ref{fig:Sustained-oscillation-from}B
and \ref{fig:Mongillo's-mean-field}. The deviations from the manifolds
will diminish with further increase of the slowness. In this limit,
the nature of the bifurcations is likely important in determining
how the system will return to the quiescent state. On the other hand,
the retention of the periodicity information in the stimulus is controlled
by the adaptive dynamics which sets the position of the driven limit
cycle of the system. This information is released after the end of
the stimulus and its manifestation is determined by the position of
this endpoint in the state space relative to the attractor manifold.
Our finding that the periodicity information is retained in the calcium
level in the WM of \cite{mongillo_synaptic_2008} is consistent with
a recent report that time perception disorders are related to WM impairment
\cite{roy_time_2012}. In fact, experiments \cite{lau_synaptic_2005}
have shown that network reverberations which is believed to be related
to WM are controlled by residual calcium level in the synapses.

We restrict the current model to the simplest dynamics of adaptation
under a periodic stimulus. We show the information of stimulus period
can be well retained with anticipative responses in accord with the
input. Such simplicity is not without limitations. For example, with
a single adaptive parameter, the return to the quiescent state will
be crossing the same region of the state space, which implies a necessary
degradation of the periodicity information before the OSR subsides.
To have an anticipative response that is transient and immune to degradation,
a more complex model, \emph{i.e.}, with more variables or involving
a bifurcation of higher codimensions \cite{izhikevich_neural_2000},
is likely required but beyond our current discussion. On the other
hand, experimental observations of such degradation in the periods
of OSR will be strongly indicative of a similar simplistic mechanism
at work.

The use of an adaptive system to fill a biological function has the
benefit of being continuously tunable and allows easier optimization
on a locally smooth landscape. Such mechanism has a lower circuitry
design cost and is better suited for recurrent environmental conditions
with varying time periods. As we have shown in this paper, the perception
of time of such systems resides in the slowest dynamics involved,
that is, the adaptive parameter $a$ in our case, or the synaptic
calcium level for a neocortical network.

This work has been supported by the Ministry of Science and Technology,
ROC under grand no. 102-2112-M-001 -009 -MY3, and National Center
for Theoretical Sciences, NCTS-SOFT/1502.

\end{document}